\documentclass[11pt,twoside]{article}
\usepackage{graphicx,epsfig,natbib,epstopdf}
\usepackage{CS18}
\begin{document}
%
% NTA: Ignore the setcounter comment here
%\setcounter{page}{7}
%
% NTA: Update your full title here
%
\title{Dynamo modeling of the Kepler F star KIC 12009504}
%
% NTA: enter your author name, affiliation/address information here
%
\author{S. Mathur$^{1}$, K. C. Augustson$^{2}$, A. S. Brun$^{3}$, R. A. Garc\'ia$^{3}$, T. S. Metcalfe$^{1}$}
\affil{$^1$Space Science Institute, 4750 Walnut Street Suite 205, Boulder, CO, USA 86001}
\affil{$^2$High Altitude Observatory, P. O. Box 3000, Boulder, CO, USA 80301}
\affil{$^3$Laboratoire AIM, CEA/DSM-CNRS-Universit\'e Paris Diderot; IRFU/SAp, Centre de Saclay, 91191 Gif-sur-Yvette Cedex, France}
\begin{abstract}
%
% NTA: Update your abstract here.
%
The {\it Kepler} mission has collected light curves for almost 4 years. The excellent quality of these data has allowed us to probe the structure and the dynamics of the stars using asteroseismology. With the length of data available, we can start to look for magnetic activity cycles. The {\it Kepler} data obtained for the F star, KIC 12009504, shows a rotation period of 9.5 days and additional variability that could be due to the magnetic activity of the star.
Here we present recent and preliminary 3D global-scale dynamo simulations of this star with the ASH and STELEM codes, capturing a substantial portion of the convection and the stable radiation zone below it. These simulations reveal a multi-year activity cycle whose length tentatively depends upon the width of the tachocline present in the simulation. Furthermore, the presence of a magnetic field and the  dynamo action taking place in the convection zone appears to help confine the tachocline, but longer simulations will be required to confirm this.
\end{abstract}
%
%
%
%
% NTA: Here is where the body of your text goes.
%
\section{Introduction}

The Kepler mission has observed 196,468 stars since its launch in April 2009. With 4-year long observations, we can study long-term variability such as rotation and magnetic activity. In addition, asteroseismic analyses of the light curves can provide us very strong constraints on the stellar fundamental parameters, structure, and dynamics \citep[e.g.][]{2014ApJS..210....1C,2011Natur.471..608B,2010Sci...329.1032G}. 
Here we present recent and preliminary 3D global-scale dynamo simulations of the Kepler target, KIC 12009504, with the ASH code, capturing a substantial portion of the convection and the stable radiation zone below it. These simulations present a multi-year activity cycle. We also notice that the presence of a magnetic field and the  dynamo action taking place in the convection zone appear to help confine the tachocline. We also performed a 2D dynamo model with the STELEM code and obtain a cycle length of ~4 years for this star.  

\section{Rotation and Magnetic activity measurement}

KIC 12009504 is a solar-like star of spectral type F with $T_{\rm eff}$=6200K. It has been continuously observed in short cadence (sampling dt=58.85s) and in long cadence (dt=29.45min) for 1440 days.
The photometric observations of KIC 12009504 were calibrated following \cite{2011MNRAS.414L...6G} to remove jumps, outliers, and instrumental trends. The flux as a function of time is shown in the top panel of Figure 1. This allows us to study more reliably the surface rotation of the star. 
To measure the surface rotation period, we performed a time-frequency analysis with the wavelets \citep[e.g.][]{1998BAMS...79...61T,2010A&A...511A..46M,2014A&A...562A.124M}. Indeed the presence of spots on the visible disk is most likely responsible for a modulation in the light curve, whose periodicity is related to the stellar rotation.

\begin{figure}[htbp]
\begin{center}
\includegraphics[width=8cm]{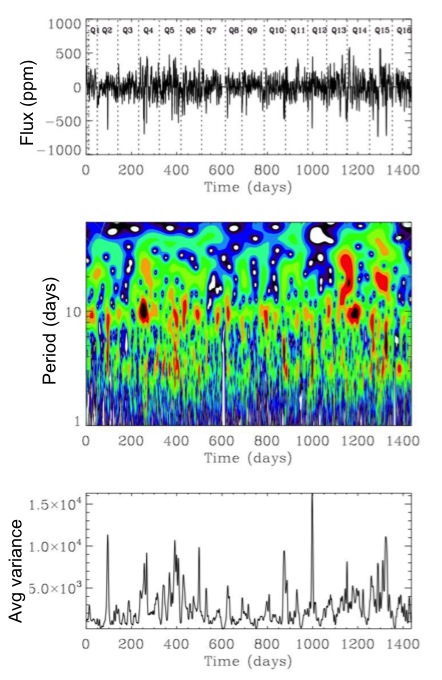}
\caption{Top panel: flux vs time of the calibrated data of KIC 12009504. Middle panel: Wavelet power spectrum as a function of time and period, where red and black colors correspond to high correlation with the Morlet wavelet frequency while green and blue correspond to low correlation. Bottom panel: magnetic proxy, i.e. projection of the wavelet power spectrum around the rotation period of the stars on the time axis \citep[from][]{2014A&A...562A.124M}.
}
\label{fig1}
\end{center}
\end{figure}

\noindent The wavelet power spectrum (middle panel of Figure~\ref{fig1}) shows the correlation  of the light curve and the Morlet wavelet as a function of time. By projecting the wavelet power spectrum on the y-axis, we observe a peak that corresponds to a surface rotation of 9.5 days  with a 99\% confidence level \citep{2014A&A...562A.124M,2014JSWSC...4A..15M,2014arXiv1403.7155G}. The width of the peak, 1.7 days, is taken as the uncertainty on the measurement.

\noindent Given the $v \sin i$ derived from spectroscopic observations ($\approx$\,8km/s) and a radius of~1.5R$_\odot$, we can determine the inclination angle of  90$\pm$27 degrees.

\noindent The photometric magnetic index computed as the standard deviation of subseries of the data with lengths 5$P_{\rm rot}$ is 131.2 ppm compared to 161 ppm for the Sun. The activity level found in KIC 12009504 is between 113 and 191 ppm, compared to the range of the Sun between 89  and 258 ppm. The range of the variability is thus smaller than the Sun's. 

\noindent By projecting the wavelet power spectrum on the time axis around the rotation period measured, we obtain a magnetic proxy (bottom panel of Figure 1). We observe a decrease in the magnetic activity until day 600, which is then followed by an increase up to the end of the observations. This suggests that if this star undergoes a magnetic activity cycle, it would have a length of at least 1400 days.

\section{Inferring the structure of KIC~12009504}

Asteroseismology is a unique tool that can allow us to directly probe the stellar interiors.
We analysed the short-cadence data and measured the global parameters (mean large frequency separation, $\langle \Delta \nu \rangle$, and frequency of maximum power, $\nu_{\rm max}$) and the individual frequencies of the acoustic modes. 
By combining the asteroseismic observables with the spectroscopic observables, we modeled the star with the Asteroseismic Modeling Portal (AMP) \citep{2009ApJ...699..373M,2014arXiv1402.3614M}.
%Figure 2 represents an \'echelle diagram obtained by taking power spectrum chunks of length $\langle \Delta \nu \rangle$ and put them on top of each other. We can see that the theoretical frequencies fit quite well the observed ones. 
Given the physics included in the stellar evolution code, the best-fit model has a mass of 1.12~M$_{\odot}$, a radius of 1.37~R$_{\odot}$, and an age of 3.64~Gyr. The depth of the convective zone is of 19\%.

\section{Modeling the Dynamo of KIC~12009504}

\subsection{The ASH code}

ASH (Anelastic Spherical Harmonic) solves the 3--D anelastic MHD equations of motion in a rotating spherical shell using a
pseudo-spectral method \citep[e.g.,][]{clune99,2004ApJ...614.1073B}.  The anelastic approximation is used to
capture the effects of density stratification without having to resolve sound waves which have short
periods ($\sim$ 5 min) relative to the decadal time-scales of solar and stellar activity.  The
solenoidality of the magnetic fields and the mass flux are enforced through a poloidal-toroidal 
streamfunction decomposition. ASH employs a mixed explicit-implicit time-stepping along with a 
pseudo-spectral algorithm for the representation of fields and the computation of their derivatives.  
All fields are decomposed horizontally into a truncated series of spherical harmonic basis functions.
Radial derivatives in ASH are computed using fourth-order finite-differences on a non-uniform radial 
grid.  Global models such as ASH are large-eddy simulations (LES) in which subgrid-scale (SGS) 
motions that cannot be resolved must be parameterized.  For the simulations carried out here, we 
have used the simple Laplacian eddy diffusion model applied to all fields.

\begin{figure}[htbp]
\begin{center}
\includegraphics[width=10cm]{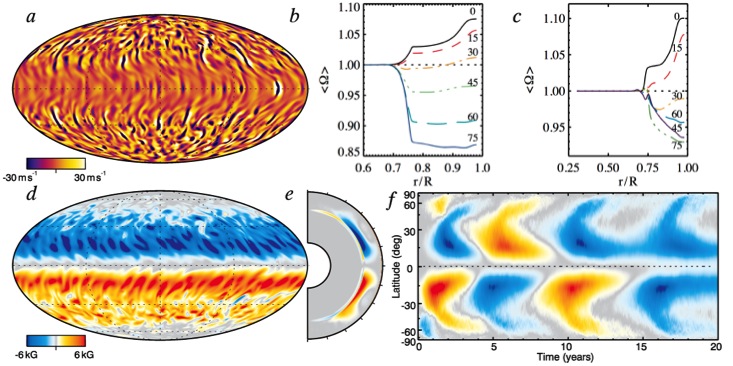}
\caption{Properties of global-scale 3D dynamo simulations.  (a) Convective patterns at mid-convection zone in a MHD simulation shown in Mollweide projection.  (b) Differential rotation in a hydrodynamic simulation shown with radius at selected latitudes.  (c) The same in a MHD simulation. (d) Toroidal magnetic field at mid-convection zone. (e) Azimuthally-averaged toroidal magnetic field shown in the meridional plane. (f) Time evolution of the toroidal field over 20 years, showing reversals and variable length cycles.}
\label{fig2}
\end{center}
\end{figure}

Using the structure of the star from the AMP best-fit model, the constraints on $P_{\rm rot}$, and the depth of the convection zone, we ran an ASH simulation encompassing the convective zone and a portion of the stable zone with r=[0.30,0.95] R where R is the stellar radius. 
We have a model resolution of 400 x 256 x 512 ($n_r$ x $n_\theta$ x $n_\phi$).
The model parameters are the viscosity ($v$), opacity ($\kappa$), diffusivity ($\eta$), which are all proportional to $\rho^{-1/2}$. These diffusivities are further reduced by a factor of $10^4$ in the radiative interior, and where Prandtl number, Pr=1/4, and the magnetic Prandtl number, Pr$_m$=1/2 are obtained throughout the domain.
The results are illustrated in Figure~\ref{fig2}.
The model possesses a latitudinal differential rotation of ~17\% (Fig. 2b), which is in agreement with uncertainty on the rotation period found above.
We find polarity reversals with a period of roughly 10 years (Fig. 2f) and an activity cycle of roughly 5 years similar to the observed trend.
Finally, we notice that oscillating magnetic field slows the spread of the tachocline relative to the hydrodynamic case (compare Fig.~2b and 2c).

\subsection{The STELEM code}

The STELEM (STellar ELEMemts) code \citep{2008A&A...483..949J} has the following characteristics:
\begin{itemize}
 \item Finite element method in space and a third order scheme in time 
  \item Domain (defined by $r_b \le r \le 1; 0 \le \theta \le \pi$) is divided into smaller regions called elements. 
  \item Boundary conditions: top as potential field and perfectly conducting bottom boundary 
  \item Adapted to solve for Babcock-Leighton unicellular flux transport mean field dynamo model
 \end{itemize}

\begin{figure}[htbp]
\begin{center}
\includegraphics[width=8cm]{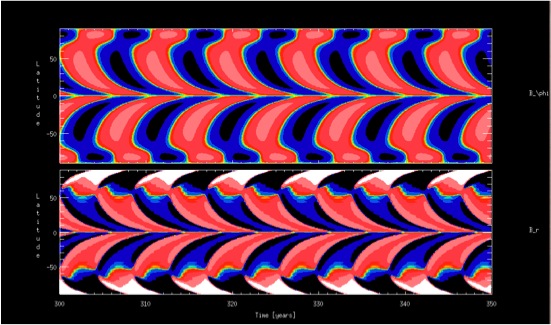}
\caption{Butterfly diagram realized with the STELEM code for a first attempt to model KIC 12009504. We see the clear periodic behavior of the dynamo, with alternance of positive (red) and negative (blue) polarities. We also note the equatorward migration of the Êmagnetic field from mid to low latitudes as well as aÊless extended polar branch.
}
\label{fig3}
\end{center}
\end{figure}

We also used the internal structure of the AMP best-fit model.
We scaled the meridional circulation amplitude up by 1.3 with respect to the solar value (20 m/s)  to take into account the higher luminosity (more vigorous convective flows). This value is consistent with ASH simulations.
Results are shown in Figure~\ref{fig3}.
We obtain an activity cycle of $\approx$4 years, which is not ruled out by the observations as they are 4-year long.

\section{Summary}
We studied the solar-like star of spectral type F, KIC 12009504, by analysing 4 years of data collected by the Kepler mission to look for signature if rotation and magnetic activty in the star. We find a rotation period of ~9.5\,$\pm$\,0.85 days.
We analysed the asteroseismic data to infer the mass and radius of the star, as well as its internal structure. The best-fit model has a convection zone depth of $\simeq$20\% of the stellar radius.
From this information, we ran preliminary 2D and 3D models of the stellar dynamo. Both models exhibit a cyclic dynamo. The internal dynamics appears to have time scales commensurable with those of the observed starpots The 3-D model further possesses large-scale intense magnetic wreaths embedded in the convective layers which undergo polarity reversals and can possibly become buoyant.

\acknowledgments{
SM acknowledges support from NASA grant NNX12AE17G. KCA acknowledges the NCAR Advanced Study Program for its support. RAG acknowledges support from FP7- IRSES and the ANR-IDEE. A.S.B. acknowledges funding by ERC 207430 STARS2 grant, ANR Toupies and INSU/PNST. NCAR is partially supported by the National Science Foundation.

}

%\bibliographystyle{aa} 
%\bibliography{/Users/Savita/Documents/BIBLIO_sav}

\end{document}